\documentclass[twocolumn,showpacs,superscriptaddress]{revtex4}

\usepackage{amsmath,amssymb}
\usepackage{graphicx}
\usepackage{dcolumn}
\usepackage{bm}
\usepackage{epstopdf}
\usepackage{amstext}
\usepackage{soul}
\usepackage[usenames, dvipsnames]{color}

\begin{document}

\title{Deterministic generation of high-dimensional entanglement between distant atomic memories via multi-photon exchange}

\author{A. Gogyan }
\email{agogyan@gmail.com} \affiliation{Institute for Physical
Research, Armenian National Academy of Sciences, Ashtarak-2, 0203,
Armenia}
\author{S. Gu\'erin}
\email{sguerin@u-bourgogne.fr} \affiliation{Laboratoire Interdisciplinaire Carnot de Bourgogne, CNRS UMR 6303, Universit\'e Bourgogne Franche-Comt\'e, BP 47870, F-21078 Dijon, France}
\author{Yu. Malakyan}
\email{yuramalakyan@gmail.com}
\affiliation{Institute for Physical Research, Armenian National
Academy of Sciences, Ashtarak-2, 0203, Armenia}

\begin{abstract}
Promising access to high-speed quantum networks relies on the creation of high-dimensional entangled memories that provide quantum communication with higher capacity of noisy quantum channels, thereby reducing the transmission time of information. Yet, the distribution of multidimensional entanglement between remote memory nodes is still faintly investigated. We propose an experimentally feasible protocol of deterministic generation of high-dimensional entanglement between distant multi-level atoms confined in high-finesse optical cavities and driven by laser pulses. Three-dimensional entanglement is generated deterministically between remote atoms by triggering a two-photon wavepacket, which mediates a superposition of states chosen by the laser pulse parameters. The efficient transfer of atomic states between remote nodes allows the construction of a three-dimensional quantum repeater, where the successful creation of entanglement can be verified by a simple method for reliable measurement of atomic ground Zeeman states.
\end{abstract}

\pacs{42.50.Ex,   03.67.Lx,   42.50.Pq,  32.80.Qk} \maketitle
\section{\protect\normalsize INTRODUCTION}

Quantum entanglement is a fundamental resource to realize quantum networks \cite{1,2} and long-distance secure key distribution \cite{3,4}. In the quantum internet \cite{5,6},  quantum repeaters (QRs) are a promising approach, enabling long-distance quantum communication over lossy channels, where the entanglement distribution between distant matter nodes is usually obstructed by the exponential loss of photons and ubiquitous decoherence over communication line. The QRs comprise local quantum memories, which are designed to store, process and release the quantum information in a reversible manner, while quantum channels link the nodes by transmitting photonic states with high transfer fidelity \cite{7,8,9,10}. A number of QR protocols have been proposed during two last decades \cite{9,10}. However, the predicted times in these schemes for the overall entanglement distribution at large distances are very long compared to the realistic quantum memory lifetimes. The communication rate is limited not only by the time required to correct the operational errors and photon losses, but mainly due to the probabilistic nature of both the entanglement generation caused by ineffective atom-photon coupling in matter nodes and entanglement swapping due to the incomplete Bell-state detection \cite{9,10,11,12,13}. As a result, a desired outcome is obtained only after many unsuccessful attempts necessitating the need for long-lived quantum memories. To date, the efficient coupling between a single photon and quantum emitters has been implemented in optical cavities \cite{14} and in one-way all-photonic QRs where the emitter spin and photons are deterministically coupled \cite{15,16,17,18,19}. Note, however, that the complexity of all-photonic structures yet poses a major obstacle to the realization of large-scale QRs. Another way to significantly improve QR efficiency, which is widely studied in recent years \cite{20,21,22,23,24,25} (for an overview see \cite{24}), is the creation of high-dimensional (HD) entangled memories and photonic states offering high data capacity and noise immunity in quantum communication. In this regard, the orbital angular momentum (OAM) states of photons provide ample opportunities \cite{23,24,25,26}, since the OAM mode-space is practically unbounded, which allows encoding information with very high degrees of freedom. Nevertheless, the reconstruction of the density matrix for high-dimensional entanglement of OAM states is difficult due to the growing amount of measured quantities as the system increases in size.

Our proposal to overcome these challenges is to construct a QR protocol that combines (i) the deterministic generation of high-dimensional entanglement over the elementary links via multi-photon exchange and (ii) near-deterministic swapping of entanglement between the entangled pairs. In this paper, we focus on the scheme of entanglement generation between two distant nodes.


The proposed scheme consists of single $F$-Zeeman-structured multilevel atoms, with $-F\le m_F\le +F$, trapped in high-finesse optical cavities. This system has an important advantage compared to the ensemble-based protocols, in which the deterministic storage and retrieval of a single photon is deteriorated by multi-photon errors. The latter are amplified, when  the probability of conversion of the atomic spin excitation into photons is increased thanks to collective interference. This imposes significant limitations on the performance of the protocols based on a single-photon detection \cite{9}.

The cavity enhanced atom-photon interaction opens a way for efficient interconversion of photonic and atomic states enabling faithful transfer of quantum states between remote atoms in a small-scale network. Here we focus on two-photon transfer, because in this case, as shown in a recent work on the deterministic transfer of photonic qutrit states between two distant atoms \cite{27}, the information is not less protected from transmission losses even over large distances than the photonic polarization qubit in conventional protocols.

\begin{figure*}[t] \rotatebox{0}{\includegraphics*
[scale =0.7]{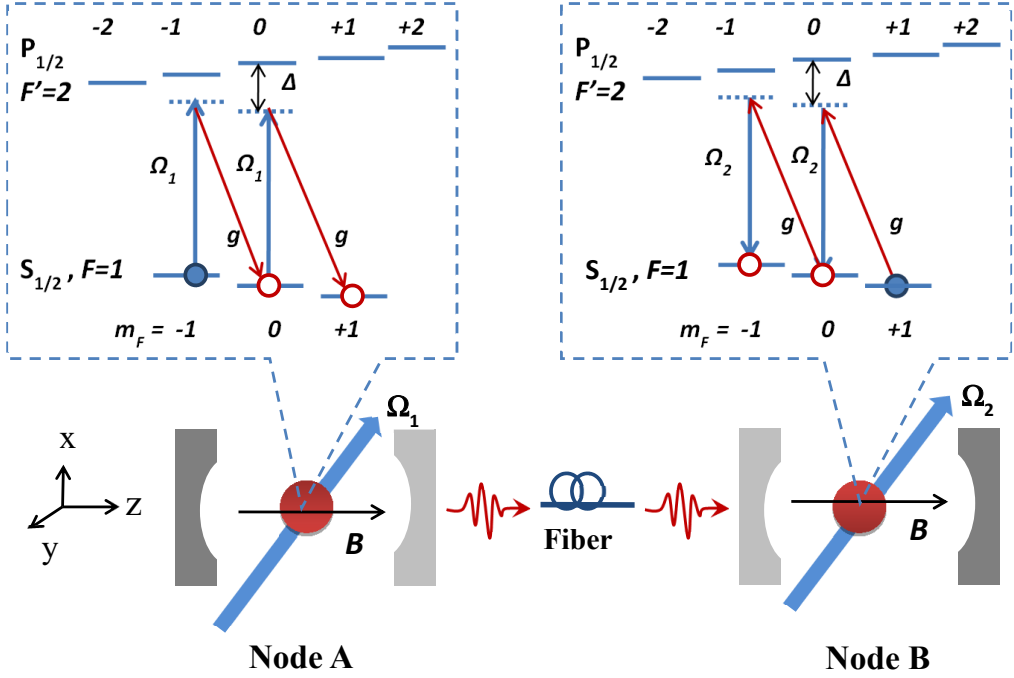}} \caption{(Color online) (a) Entanglement generation between remote atoms via two-photon exchange. The insets show the atomic level structure and interaction of the atoms with laser fields $\Omega_{1,2}$ and external magnetic field B applied along the cavity axis. The sequence of $\sigma^-$-polarized cavity photons (thin red lines) are generated in the left and absorbed in the right cavities. The initial populations of atomic states are shown by filled circles \label{fig:fig1}}
\end{figure*}

For alkali atoms, the corresponding configuration of atom-photon interaction is shown in Fig. \ref{fig:fig1} for the $S_{1/2}(F = 1) \rightarrow P_{1/2}(F' = 2)$ transition, where the upper states $F' = 2$ and $F' = 1$ are well separated. In the sending node \textit{A}, the atom is initially prepared, e.g., in the ground state $m_F=-1$ (Fig. \ref{fig:fig1}, left) and interacts with a linearly polarized control laser pulse $\Omega_1(t)$ and a $\sigma^-$-polarized cavity mode in Raman configuration with intermediate state $F'=2$. Hereafter, all physical quantities in the first and second cavities are labeled with indices 1 and 2, respectively. This interaction coherently distributes the atomic population over the Zeeman sublevels $m_F = -1,0,1$ with different amplitudes, meanwhile generating two photons in the cavity. Atomic population amplitudes and photon temporal profiles are deterministically controlled by the laser  pulse under realistic high signal-to-noise ratio. An external magnetic field applied parallel to the cavity axis removes the Zeeman degeneracy to finely tune the $\sigma^-$-polarized cavity photons to the two-photon Raman resonance on the $m_F\rightarrow m_{F}+1$ transition, as well as to quench the decoherence of the atomic superposition state by ambient magnetic fields. The cavity photons are efficiently coupled to a low-loss single-mode optical fiber as a multi-photon
wave packet. The fiber transmits it to the receiving system \textit{B }(Fig. \ref{fig:fig1}, right), where a single atom, initialized in the state $m_F=+1$, interacts with a linearly polarized $\Omega_2(t)$ laser pulse. It induces a controlled reversed absorption of the incoming photons with unit probability, thus creating a three-dimensional entanglement between three-level ground states of the sending and receiving atoms. The shape and intensity of the $\Omega_2(t)$ laser pulse is derived from the key requirement to entirely exclude the photon leakage from the second cavity indicating that the quantum output field is zero at all times. Importantly, the absorption amplitudes of the photons in the receiving node are not sensitive to their spectral properties due to the integral dependence on the photon temporal profiles (see Sec. IIIA). Under these conditions, the  entanglement generation between nodes is much faster than the decoherence time of  entanglement. To the best of our knowledge, this proposal is the first protocol for entanglement creation between atomic qutrits, which can be directly extended to higher dimensional $(d>3)$ atomic states using the results of Ref.\cite{31}.

Different from the probabilistic protocols \cite{9}, the entanglement between two distant memories is implemented in our scheme by direct transfer of photons, and does not arise in a "heralded" way. At the same time, the heralding of the successful entanglement makes the process resistant to operational errors such as fluctuating photon arrival times and detector inefficiency \cite{9,12,13,28,29}. However, the implementation of this approach in our case of multiphoton absorption is not yet available. Instead, to verify the successful storage of incoming photons into the second node we propose a simple method for reliably measuring the ground state of the receiving atom.

This paper is organized as follows. In the next section, we present the interaction setup and solve the evolution equations for the atomic state amplitudes and the cavity field. The necessary conditions for deterministic generation of $\sigma^-$-polarized cavity photons in the sending node A are derived. We calculate the populations of ground Zeeman states of the atom for different durations of the Gaussian pumping laser and obtain the temporal profiles of emitted photons. These data are important to control the entanglement between the atoms. Here we find also the final state of quantum field produced from the node A. In Sec. III, the complete absorption of incoming photons at the node B is calculated under strongly prohibiting the photon loss from the second cavity. We demonstrate the creation of three-dimensional entanglement between the atoms with maximally entangled two-qutrit state. Here we also present a new scheme for measuring the atomic ground states, which is designed for testing the robustness of our scheme against the noisy local operations and photon losses in the quantum channels. We summarize the results in Sec. IV.

\section{\protect\normalsize DETERMINISTIC PRODUCTION OF TWO-PHOTON PROPAGATING STATE FROM THE SENDING CAVITY}

\subsection{Model}
To describe the production of the cavity photons in the sending node A in Fig. \ref{fig:fig1}, we use the method for deterministic generation of a stream of multiphoton pulses in a single atom-cavity system proposed in \cite{30,31}. The off-resonant Raman scattering of the linearly polarized $\Omega_1$ laser field generates $\sigma^{-}-$polarized cavity photons on the transitions $|F'=2, m_{F'} = m_F\rangle_1 \rightarrow |F=1, m_F+1\rangle_1$ with the coupling strength $g$, thus transferring the atom into the next ground-state Zeeman sublevel $m_F+1$. The quantization axis is taken parallel to the cavity axis and to the external magnetic field, which is chosen strong enough to prevent the generation of $\sigma^+-$polarized cavity photons due to their off-resonant interaction with the atom. The one-photon detunings  $\Delta$, taken the same for both atoms in the nodes \textit{A} and \textit{B}, are chosen (i) small enough compared to the upper level hyperfine splitting, but (ii) much larger than the cavity decay rate $k$, the natural spontaneous decay rate $\gamma_{\text{sp}}$ of the atom and the Rabi and Larmor frequencies: $|\Delta| \gg k, \gamma_{\text{sp}}, \Omega_{1}(t), |\Delta_B^{(F, F')}|$. Here $\Delta_B ^{(F,F')}=g_L ^{(F,F')}\mu_B B/\hbar$ is the Zeeman splitting of the ground and excited states in the magnetic field $B$, with $g_L^{(F, F')}$ the Land\'e factor and $\mu_B$ the Bohr magneton. These conditions allow one to neglect the spontaneous losses from upper levels and dephasing effects induced by other excited states. In this far off-resonant case and for slowly varying laser fields: $d\Omega_1(t) / dt \ll \Delta\Omega_1(t)$, one can adiabatically eliminate the upper atomic states and define the effective Raman atom-photon coupling
\begin{equation}
G_1=g\Omega_1/\Delta.\label{eq:1}
\end{equation}
We describe the laser pulse as
\begin{equation}
E_1(t)={\mathcal E}_1f_1^{1/2}(t)\exp (-i\omega t)+h.c.,\label{eq:2}
\end{equation}
where $f_1(t)$ features its temporal profile of duration $T_1$ and ${\mathcal E}_1$ is the peak amplitude defining the one-photon detuning $\Delta=\omega_{FF'} - \omega$ and the peak Rabi frequency $\Omega_1=\mu_{F,F'}{\mathcal E}_1/\hbar$. For simplicity, we assume here that the dipole matrix elements $\mu _{F,F'}$ of the $|F', m_{F'}\rangle\rightarrow |F, m_F\rangle$ transitions do not depend on the magnetic quantum numbers. This actual dependence can be easily taken into account in the numerical simulations as is shown in Ref. \cite{31}. Hereafter  the phase of ${\mathcal E}_1$, serving as the phase reference, is taken zero without loss of generality.

The process of producing photons from the sending node is deterministic if it can be controlled by the laser pulse $\Omega_1(t)$, for which two basic requirements must be met. First, the generated photons must leave the cavity before they can be reabsorbed by laser-stimulated Raman scattering on $|F=1, m_F+1\rangle_1 \rightarrow |F=1, m_F\rangle_1$ transitions, that is, the two-photon Rabi oscillations with frequency $G_1$ must be slower than the rate of photon leakage from the cavity
\begin{equation}
|G_1|< k.\label{eq:3}
\end{equation}
Secondly, all relaxation processes including the optical pumping from $|F,m_F\rangle_1$ into states $|F,m_{F}\pm1\rangle_1$, and the loss of atomic populations due to spontaneous decay of the upper state $|F'=2\rangle_1$ into the states outside the system should be negligible compared to the cavity photon generation rate $\alpha_1=4G_1^2/k$, that is $\alpha_1f_1(t)\gg\Gamma_1(t)$, where $\Gamma_1(t)=\frac{\Omega_1^2}{\Delta^2}f_1(t)\gamma_{\text{sp}}$ is the total spontaneous decay rate induced by the $\Omega_1(t)$ pump pulse \cite{31}. This defines the signal-to-noise ratio or the single-atom-cavity cooperativity $C_c$ \cite{32}
\begin{equation}
R_{\text{sn}} =C_c= \frac{\alpha_1 f_1(t)}{\Gamma_{1}(t)} = \frac{4g^2}{ k\gamma_{\text{sp}}} \gg 1.\label{eq:4}
\end{equation}
This condition is clearly fulfilled in high-finesse optical cavities with $g > k, \gamma_{\text{sp}}$.

After adiabatic elimination of the upper states, the effective interaction Hamiltonian in RWA takes the form
\begin{align}
H&=\hbar \Biggl[\sum_{m_F=-F}^{F}\biggl ( \frac{g^2}{\Delta}a_1^{\dag}a_1+f_1(t)\frac{\Omega_1^2}{\Delta}\biggr )\sigma_{m_F,m_F} \nonumber \\
&+ G_1 f_1^{1/2}(t) \sum_{m_F=-F}^{F-1}\biggl (a_1^{\dag}(t) \sigma_{m_F+1,m_F}(t)+h.c.\biggr )\Biggr],\label{eq:5}
\end{align}
where $\sigma_{i,j}(t) = |i \rangle_1\langle j |_1$ and $a_1(t),a_1^\dag(t)$ are the atomic and photonic mode operators in the first cavity, respectively. The two first terms in Eq. (5) describe the Stark shifts of atomic ground states induced by the cavity field and laser pulse $\Omega_1(t)$, respectively. With the adopted approximation of equal dipole moments, the Stark shifts induced by the laser field have no influence on the photon  generation, while the Stark shift $g^2/\Delta$ is the same for all transitions and can be included into the cavity mode frequency: $\omega_c \rightarrow \omega_c-g^2/\Delta$. With the adopted approximation of equal dipole moments, the Stark shifts induced by the laser field have no influence on the photon generation, since their difference between two neighboring atomic ground states is zero for all transitions $m_F\rightarrow m_F+1$, while the Stark shift $g^2/\Delta$ can be included into the cavity mode frequency. 
In real atoms, where the dependence of $g$ and $\Omega_1$ on Clebsch-Gordan coefficients leads to non-vanishing Stark-shift difference, the latter can be made negligibly small as compared to the cavity decay rate $k$ by appropriately choosing the system parameters \cite{31}, [Sec. IIIC].

\subsection{Evolution of output photonic state}
The elementary link in our QR protocol is a cascade system consisting of two consecutive cavities in a chain, where the subsequent cavity is driven by photons coming from the previous cavity \cite{33,34} without photon reflection. Our goal is to absorb the incoming photons in the second cavity with a unit probability, which requires precise knowledge of the imperfect control of the wave functions of emitted photons by the $\Omega_1(t)$ laser in order to correct operational errors, primarily fluctuations in the time-of-arrival of photons, as indicated in Introduction.

Here we obtain the output state of the emitted photons from their total flux defined in units of photons per unit time by
\begin{equation}
\frac{dn_{\text{out}}(t)}{dt} = \langle a^{\dag}_{1,\text{out}}(t)a_{1,\text{out}}(t)\rangle,\label{eq:6}
\end{equation}
where $n_{\text{out}}(t)$ is the mean number of photons emitted from the first cavity in the time interval $(-\infty,t]$. The output field $a_{1,\text{out}}(t)$ is connected to the input $a_{1,\text{in}}(t)$ and cavity mode $a_1(t)$ operators by the input-output relation $a_{1,\text{out}}(t)-a_{1,\text{in}}(t)=\sqrt{k}a_1(t)$ \cite{35}, and $a_1(t)$ is obtained through the atomic operators from the Heisenberg-Langevin equation along with the Hamiltonian (5) and in the adiabatic limit $kT_1\gg 1$ as \cite{31}
\begin{equation}
a_1(t)=-\frac{2i G_1}{k} f_1^{1/2}(t)\sum_{m_F=-F}^{F-1}\sigma_{m_F+1,m_F}(t) - \frac{2}{\sqrt{k}}a_{1,\text{in}}(t).\label{eq:7}
\end{equation}
In the sending node, the input field $a_{1,\text{in}}(t)$ is in the vacuum state $\langle a^{\dag}_{1,\text{in}}(t) a_{1,\text{in}}(t)\rangle = 0$ and will be ignored in further calculations.

Then, combining Eqs. (6) and (7), we have
\begin{equation}
\frac{dn_{\text{out}}(t)}{dt}=\alpha_1f_1(t)\sum_{m_F=-1}^{0}\langle \sigma_{m_F}(t)\rangle,\label{eq:8}
\end{equation}
where the equations for the Zeeman sublevel populations $\langle\sigma_{m_F}(t)\rangle \equiv \langle\sigma_{m_F,m_F}(t)\rangle$ with  arbitrary $F$ were derived in \cite{31} from the master equation for the whole density matrix of the system. For $F=1$, they are solved from the initial conditions $\langle \sigma_{-1}(-\infty)\rangle =1$, $\langle \sigma_0(-\infty)\rangle =\langle \sigma_1(-\infty)\rangle=0$. Using condition (4) to neglect the spontaneous losses, we get
\begin{subequations}
\begin{eqnarray}
& & \langle \sigma_{-1}(t)\rangle= e^{-\vartheta(t)},\label{eq:9a}\\
& & \langle \sigma_0(t)\rangle= \vartheta(t)e^{-\vartheta(t)},\label{eq:9b}\\
& & \langle \sigma_1(t)\rangle=1-\sum_{m_F=-1}^{0}\langle\sigma_{m_F}(t)\rangle,\label{eq:9c}
\end{eqnarray}
\end{subequations}
where the variable
\begin{equation}
\vartheta(t) = \alpha_1\int \limits_{-\infty}^tf_1(t')dt'\label{eq:10}
\end{equation}
is proportional to the $\Omega_1(t)$ pulse energy confined in the $(-\infty,t]$ interval. Meanwhile, in the absence of Rabi oscillations,  the atomic coherence is always zero $\langle\sigma_{m_F,m_F^\prime}(t)\rangle\equiv0, m_F\neq m_F^\prime$.

The wave functions $\Phi_{\text{I,II}}(t)$ of the two emitted photons can be derived from the total flux (8) as a sum of the first and second photon fluxes, proportional to the atomic populations $\langle \sigma_{-1}(t)\rangle$ and $\langle \sigma_0(t)\rangle$, respectively:
\begin{subequations}
\label{eq:11}
\begin{eqnarray}
\frac{dn_{\text{I}}(t)}{dt} &=& |\Phi_{\text{I}}(t)|^2= \alpha_1f_1(t)e^{-\vartheta(t)},\label{eq:11a}\\
\frac{dn_{\text{II}}(t)}{dt} &=& |\Phi_{\text{II}}(t)|^2= \alpha_1f_1(t)\vartheta(t)e^{-\vartheta(t)},\label{eq:11b}
\end{eqnarray}
\end{subequations}
where $\Phi_{\text{I,II}}(t)$ having the phase of control field $\Omega_1(t)$ are real.

Having thus obtained the photon wave functions, it remains to construct the final state of the sending cavity and outgoing photons. In the model we consider, the basis states of the sending system at asymptotic times $t>k^{-1}$ are the tensor product state of (i) atomic ground Zeeman states $|m_F\rangle$ with (ii) photonic states $|j\rangle$ corresponding to free propagating $\sigma^-$-polarized $j$ photons of frequency $\omega_c$, which are coupled to a single-mode optical fiber, and (iii) the vacuum cavity mode state $|0\rangle_{1c}$.
The final state $|\Psi_{1,\text{fin}}(t)\rangle$ can be then expanded in this basis as an entangled state between the atom and the outgoing light, of the form
\begin{equation}
|\Psi_{1,\text{fin}}\rangle = \biggl(\sum_{m_F=-1}^{1}\beta_{m_F,j} |m_F\rangle_{1}|j=m_F+1\rangle_{ph}\biggr )|0\rangle_c,\label{eq:12}
\end{equation}
where the real coefficients $\beta_{m_F,j}$ are probability amplitudes of the atom asymptotically in the state $|m_F\rangle_1$ corresponding to the photon number $j=m_F+1$. Since the second index $j$ of $\beta_{m_F,j}$ is uniquely determined by $m_F$, we will safely omit it below.

The photonic states $|j=1\rangle_{ph} = |1_{\Phi_\text{I}}\rangle|0_{\Phi_\text{II}}\rangle$ and $|j=2\rangle_{ph} = |1_{\Phi_\text{I}}\rangle|1_{\Phi_\text{II}}\rangle$ can be defined as the product of single-photon states with wave functions $\Phi_\text{I},\Phi_\text{II}$ introduced in Eqs. (11). These states are defined as $b_i^{\dagger}|0_{\Phi_i}\rangle=|1_{\Phi_i}\rangle$, where the independent operators $b_i,i=I,II$ have the standard boson commutation relations $[b_i,b^\dagger_k]=\delta_{ik}$. In this case, the operator $a_{1,\text{out}}(t)$ is expressed through $b_i$ as \cite{36}
\begin{equation}
a_{1,\text{out}}(t) = \sum_{i=\text{I}}^{\text{II}}\Phi_i(t)b_i,
\end{equation}\label{eq:13}
which gives
\begin{equation}
a_{1,\text{out}}(t)|n_{\Phi_\text{I}}\rangle|n_{\Phi_\text{II}}\rangle = \sum_{i\neq k}n_{\Phi_i}^{1/2}\Phi_i(t)|n_{\Phi_i}-1\rangle|n_{\Phi_k}\rangle. \label{eq:14}
\end{equation}
Using the relations $\beta_{m_F}^2 = \langle \sigma_{m_F}(\infty)\rangle$, followed from Eq. (12), and Eqs. (9), $\beta_{m_F}$ are found to be
\begin{subequations}
\begin{eqnarray}
& & \beta_{m_F=-1,j=0}=e^{-\vartheta(\infty)/2},\label{eq:15a}\\
& & \beta_{m_F=0,j=1}=[\vartheta(\infty) e^{-\vartheta(\infty)}]^{1/2},\label{eq:15b}\\
& & \beta_{m_F=1,j=2}=[1-(1+\vartheta(\infty))e^{-\vartheta(\infty)}]^{1/2}.\label{eq:15c}
\end{eqnarray}\label{eq:15}
\end{subequations}
We show in the next section that these coefficients determine the amount of entanglement between the sending and receiving atoms. The dynamics of the atomic populations $\langle \sigma_{m_F}(t)\rangle$ is shown in Fig. 2 for three durations $T_1$ of a Gaussian $\Omega_1(t)$ laser pulse $f_1(t)=e^{-(t/T_1)^2}$, where the realistic parameters $(g, k, \gamma_{\text{sp}}, \Omega_1, \Delta^{F}_B, \Delta^{F'}_B, \Delta) = 2\pi\times(12, 3, 5.87, 7, -12, 4, 100)$MHz \cite{37,38} were used to ensure a high cooperativity in Eq.(4): $C_c \sim 30$. We list the numerical results for $\beta_{m_F}^2$ in Table 1. For the case of $T_1=0.12\mu s$, the photon wave functions $\Phi_{\text{I,II}}(t)$, which are real and positive, are shown in Fig. 3. We observe that the photons have different pulse areas, which can be an issue for creating the maximal entanglement between the atoms. This is discussed in Sec. IIIA.
\begin{figure}[t] \rotatebox{0}{\includegraphics*
[scale =0.7]{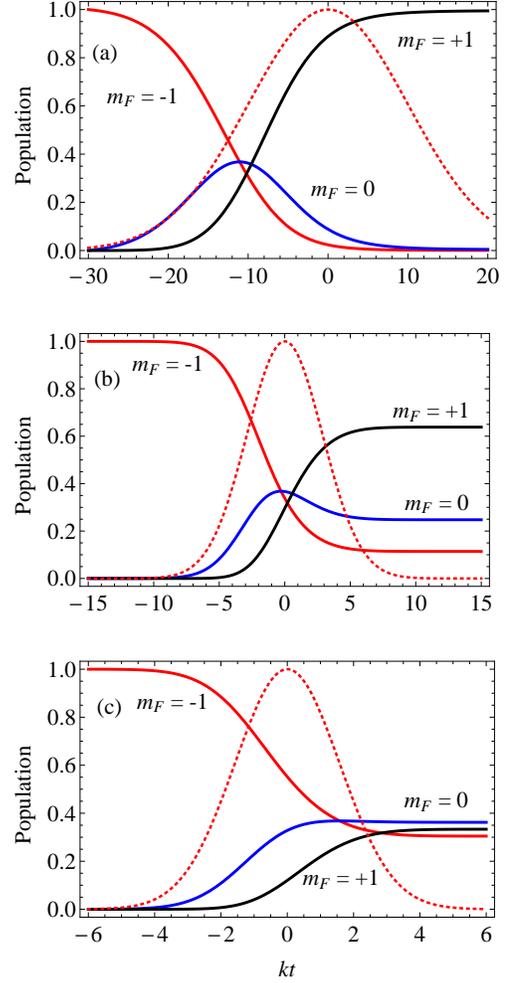}} \caption{(Color online) Evolution of atomic populations of the ground Zeeman states $|m_F=-1\rangle_1$ (red), $|m_F=0\rangle_1$ (blue) and $|m_F=+1\rangle_1$ (black) for three durations $T_1=0.75\mu$s (a); $0.22\mu$s (b) and $0.12\mu$s (c) of  Gaussian laser pulse $f_1(t)=e^{-t^2/T_1^2}$ (dashed line). The atom is initially prepared in the states $|F=1,m_F=-1\rangle_1$. For the parameters see the text.
\label{fig:fig2}}
\end{figure}

\begin{figure}[t] \rotatebox{0}{\includegraphics*
[scale =0.8]{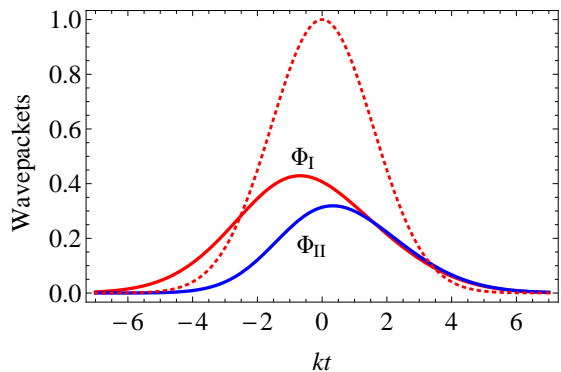}} \caption{(Color online) Wave-functions of the photons (indicated on the curves) corresponding to the dynamics shown in Fig. 2 c.
\label{fig:fig3}}
\end{figure}

\section{\protect\normalsize ENTANGLEMENT GENERATION VIA ELEMENTARY LINKS}
For the system of cascade memories in Fig. 1, assuming no losses in the communication channel, the final state Eq. (12) serves as an input state for the receiving node, where the atom with the same level configuration is initially prepared in the state $|F=1,\bar{m}_F=1\rangle_2$. Thus, the initial state of the receiving system is given by
\begin{equation}
|\Psi_{2,\text{in}}\rangle = |\bar{m}_F=1\rangle_{2}\Bigl[\sum_{m_F=-1}^{1}\beta_{m_F} |m_F\rangle_{1}|j=m_F+1\rangle_{ph}\Bigr].\label{eq:16}
\end{equation}

The atom interacts on the transition $|F=1,\bar{m}_F=1\rangle_2\rightarrow|F'=2,\bar{m}_F'=1\rangle_2$ with the linearly polarized laser pulse of frequency $\omega$, duration $T_2$ and the peak Rabi frequency $\Omega_2$, which can be different from $\Omega_1$. Similarly, the same magnetic field as in the sending node is applied to provide two-photon resonant Raman transition with $\sigma^-$ polarized cavity photons.

The key requirement for complete storage of incoming photons in the receiving node is to prevent reflection or leakage of photons from the second cavity, which implies that the output quantum field is always zero. This amounts to formally setting
\begin{equation}
a_{2,\text{out}}(t)=0\label{eq:17}
\end{equation}
in the input-output relation $a_{2,\text{out}}(t)-a_{2,\text{in}}(t)=\sqrt{k}a_2(t)$ for the second cavity. Then, using $a_{2,\text{in}}(t)=a_{1,\text{out}}(t-\tau)$, we have
\begin{equation}
a_2(t) = -\frac{1}{\sqrt{k}}a_{1,\text{out}}(t-\tau),\label{eq:18}
\end{equation}
where $\tau= L/c$ with \textit{L} the distance between the nodes is the time needed for the photons to travel from one cavity to the next. In the theory of cascade systems, which is also our scheme, it is assumed that the dynamics of the receiving node does not depend on \textit{L} \cite{33,34}. However, this is true only if the path length fluctuations change $\tau$  insignificantly. Due to these fluctuations the incoming photons undergo the phase noise during the travel time. In a long fiber, the phase noise has been measured over the scale of photon travel time in \cite{39} using Mach-Zehnder interferometry. The results show that the phase over 36 km long interferometer remains stable at a level of 0.1rad for the duration of around 100$\mu$s, which corresponds to a path length fluctuation of several tens of nm for a wavelength 1550nm and even less for visible light. Obviously, this induces a negligible change in the delay time, which justifies the above assumption. This allows safely omitting $\tau$ in all equations.

\begin{table}[]
\caption{Populations of the sending atom in Zeeman states $|F=1,m_F=-1,0,1\rangle_1$ for three durations $T_1$. The last column shows the amount of entanglement between the atoms.}
\begin{tabular}{lllll}
\hline
\hline
 $T_1 (\mu$s) \ \ \ \ \ & $\beta_{-1}^2$ \ \ \ \ \ \ \ \ \ \ \ & $\beta_{0}^2$ \ \ \ \ \ \ \ \ \ \ \ & $\beta_{1}^2$ \ \ \ \ \ \ \ \ \ \ \ & $E$ \ \ \ \  \\
 0.75  & 0.00065 & 0.0048  & 0.995 & 0.051 \\
 0.22  & 0.11    & 0.25    & 0.64  & 1.27\\
 0.12  & 0.31    & 0.36    & 0.33  & 1.58\\ \hline
\end{tabular}
\end{table}

\subsection{Photon storage in the receiving node. Maximally entangled state of two atoms}
The dynamics of photon absorption in the receiving system is governed by an effective interaction Hamiltonian in the form
\begin{equation}
H=\hbar f_2^{1/2}(t) \bigl [G_2^*(\varrho_{0,1}+\varrho_{-1,0})a_2(t)+G_2a_2^{\dag}(t)(\varrho_{1,0}+\varrho_{0,-1})\bigr]\label{eq:19}
\end{equation}
where $\varrho_{i,j} = |\bar{m}_F=i\rangle_2\langle \bar{m}_F=j|_2$ are the operators of the receiving atom, $G_2=g\Omega_2/\Delta$ and $f_2(t)$ represents the time profile of the pumping laser in the second cavity.

In general, the state of the receiving system can be expanded as
\begin{eqnarray}
|\Psi_2(t)\rangle=\sum_{m_F=-1}^{1}\beta_{m_F}|m_F\rangle_1 \nonumber\\
\times\sum_{\bar{m}_F=-m_F}^1\gamma_{\bar{m}_F,j}(t)|\bar{m}_F\rangle_2|j=\bar{m}_F+m_F\rangle_{ph},\label{eq:20}
\end{eqnarray}
where $\gamma_{\bar{m}_F,j}(t)$ are the amplitudes of the probabilities that at time $t$ the receiving atom is in the state $|\bar{m}_F\rangle_2$ and the number of incoming photons is $j=\bar{m}_F+m_F$ provided that the first atom is in the state $|m_F\rangle_1$. They satisfy the normalization equation
\begin{equation}
\sum_{m_F=-1}^{1}\sum_{\bar{m}_F=-m_F}^1\beta_{m_F}^2|\gamma_{\bar{m}_F,\bar{m}_F+m_F}(t)|^2=1.\label{eq:21}
\end{equation}
The structure of Eq. (20) is the result of the double action of the first term of Hamiltonian (19) on the initial state (16) and shows that the photons are partially entangled with both atoms, which is evidenced by the dependence of the photonic state on magnetic numbers of two atoms. Initially, the photons are entangled with the first atom shown in Eq. (16), but not with the second one. However, under the conditions derived below, the photons are completely absorbed by the second atom leading to the conversion of the atom-photon entanglement into the entanglement between the two atoms with maximum value at $\beta_j=1/\sqrt{3}$. The entire dynamics of this effect is captured by the evolution of the coefficients $\gamma_{\bar {m}_F,j}(t)$, as we show further.

We find $\gamma_{\bar{m}_F,j}(t)$ from the Schr\"odinger equation for $|\Psi_2(t)\rangle$ with the use of Hamiltonian (19), where the operator $a_2(t)$ is replaced by $a_{1,\text{out}}(t)$ from Eq. (18) with $\tau=0$. We recognize that the action of the latter on the two-photon state $|1_{\Phi_\text{I}},1_{\Phi_\text{II}}\rangle$ generates, according to Eq. (13), the superposition of two one-photon states
\begin{equation}
a_{1,\text{out}}(t)|1_{\Phi_\text{I}},1_{\Phi_\text{II}}\rangle=\frac{1}{\sqrt{2}}\sum_{i=+,-}\Phi_i(t)|i\rangle,\label{eq:22}
\end{equation}
which arises in the equation for $\gamma_{\bar{m}_F=0,j=1}(t)$. Here $\Phi_{\pm}(t)=\Phi_{I}(t)\pm\Phi_{II}(t)$ and $|\pm\rangle=\frac{1}{\sqrt{2}}(|1_{\Phi_I},0_{\Phi_{II}}\rangle\pm|0_{\Phi_I},1_{\Phi_{II}}\rangle)$. A key approximation to achieve the complete absorption of the photons consists in considering
\begin{equation}
\eta(t)\sim\zeta(t), \label{eq:23}
\end{equation}
with the variables
\begin{subequations}
\begin{eqnarray}
& & \eta(t) = 2\frac{|G_2|}{\sqrt{k}}\int \limits_{-\infty}^tf_2^{1/2}(t')\Phi_I(t')dt',\label{eq:24a}\\
& & \zeta(t) = \frac{|G_2|}{\sqrt{k}}\int \limits_{-\infty}^tf_2^{1/2}(t')\Phi_+(t')dt',\label{eq:24b}
\end{eqnarray}
\end{subequations}
featuring the time integral of the photon fluxes multiplied by the profile of the second laser driving the receiving node.

How to satisfy this approximation is discussed below.
This implies taking only the leading term $~\Phi_{+}(t)$ into account in Eq. (22) while integrating the Schr\"odinger equation. This gives a simple form for the equations of $\gamma_{\bar{m}_F,j}(t)$:
\begin{subequations}
\begin{eqnarray}
\frac{d\gamma_{ 0,0}(t)}{d\eta} = \frac{i}{2}\gamma_{1,1}(t)e^{-i\varphi_2},\label{eq:25a}\\
\frac{d\gamma_{1,1}(t)}{d\eta} = \frac{i}{2}\gamma_{0,0}(t)e^{i\varphi_2},\label{eq:25b}\\
\frac{d\gamma_{1,0}(t)}{d\eta} = 0, \label{eq:25c}
\end{eqnarray}\label{eq:25}
\end{subequations}
and
\begin{subequations}
\begin{eqnarray}
\frac{d\gamma_{-1,0}(t)}{d\zeta} = \frac{i}{\sqrt{2}}\gamma_{0,1}(t)e^{-i\varphi_2},\label{eq:26a}\\
\frac{d\gamma_{0,1}(t)}{d\zeta} = \frac{i}{\sqrt{2}}\biggl(\gamma_{1,2}(t)e^{-i\varphi_2}+\gamma_{-1,0}(t)e^{i\varphi_2}\biggr),\label{eq:26b}\\
\frac{d\gamma_{ 1,2}(t)}{d\zeta} = \frac{i}{\sqrt{2}}\gamma_{0,1}(t)e^{i\varphi_2},\label{eq:26c}
\end{eqnarray}\label{eq:26}
\end{subequations}
where $\varphi_2$ is the phase of $\Omega_2$ field.

These equations are solved with the initial conditions $\gamma_{1,i}(-\infty)=1, i=0,1,2,$ and zero for the remaining coefficients, which are obtained by comparing Eq. (20) with the initial state (16). For $\varphi_2=\pi/2$, we get
\begin{equation}
\gamma_{1,0}(t)=1,\quad\gamma_{0,0}(t) = \sin[\eta(t)/2], \quad \gamma_{1,1}(t) = \cos[\eta(t)/2],
\end{equation}\label{eq:27}
and
\begin{subequations}
\begin{eqnarray}
\gamma_{1,2}(t) = \frac{1}{2}\biggl [1+\cos\zeta(t)\biggr],\label{eq:28a}\\
\gamma_{0,1}(t)=\frac{1}{\sqrt{2}}\sin\zeta(t),\label{eq:28b}\\
\gamma_{-1,0}(t) = \frac{1}{2}\biggl [1-\cos\zeta(t)\biggr],\label{eq:28c}
\end{eqnarray}\label{eq:28}
\end{subequations}
which satisfy Eq. (21).

It is evident that the complete absorption of photons in the second cavity is achieved, when the time profile of the second laser is such that both integrals (24) are ultimately $\pi$:
\begin{equation}
\eta(\infty)=\zeta(\infty)=\pi,
\label{eq:29}
\end{equation}
because in this case only the amplitudes $\gamma_{i,0}(\infty)=1, i=1,0,-1,$ remain different from zero, indicating that, at large times $t>T_2$, the quantum field is in the state $|j=0\rangle_{ph}$ with zero photon number, while the second atom occupies the Zeeman states with $\bar{m}_F=-m_F$.

The described process is two consecutive $\pi$-pulse excitations of two Raman transitions $|F=1,\bar{m}_F=1\rangle_2\rightarrow |F=1,\bar{m}_F=0\rangle_2$ and $|F=1,\bar{m}_F=0\rangle_2\rightarrow |F=1,\bar{m}_F=-1\rangle_2$ with effective (two-photon) pulse areas $\eta(\infty)$ and $\zeta(\infty)$, respectively, which ensure the photon storage into the ground states of the second atom with unit probability.

Under condition (29), two atoms ultimately turn out to be in a pure entangled state in the three dimensional space of atomic states
\begin{eqnarray}
|\Psi_{fin}\rangle_{12}=|\Psi_2(\infty)\rangle=\sum_{m_F=-1}^{1}\beta_{m_F}|m_F\rangle_1 |-m_F\rangle_2.
\end{eqnarray}\label{eq:29pr}
The entanglement between the atoms, characterized by the quantity $E=-\sum_{m_F=-1}^{1}\beta_{m_F}^2\log_2\beta_{m_F}^2$ \cite{40}, is maximal for equal $\beta_{m_F}=\frac{1}{\sqrt{3}}$. This case leading to $E_{\max}\simeq 1.58$ is realized with a high accuracy for $T_1=0.12 \mu s$ as shown in Fig. 2c. The corresponding values of $\beta_{m_F}$ are given in Table 1.

\begin{figure}[t] \rotatebox{0}{\includegraphics*
[scale =0.8]{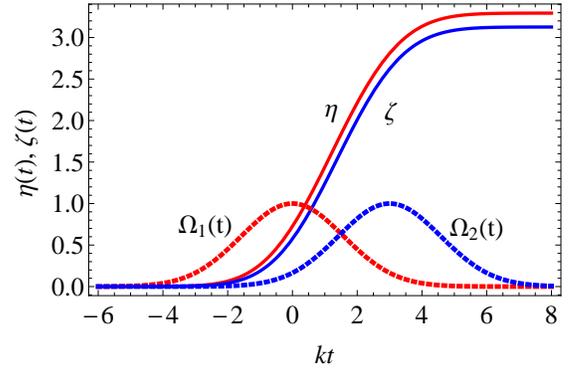}} \caption{(Color online) Variables $\eta(t)$ (red) and $\zeta(t)$ (blue) as functions of time for $\Omega_2=4\Omega_1$ and $T_1=T_2=0.12\mu$s. The profiles of $\Omega_{1,2}(t)$ laser fields are shown by red and blue dashed line, respectively.
\label{fig:fig4}}
\end{figure}

The robustness of our results is defined only by the stability of laser pulse energy, because as follows from Eqs.(10), (15), (24) and (27,28), the coefficients $\beta_i$ and $\gamma_{ij}$, and hence, the entangled state of the first atom and the outgoing photons, Eq.(12), as well as the final entanglement between two remote atoms, Eq.(30), depend only on the total energy of the pump pulses. The energy stability factor [(max - min)/mean] of modern nanosecond lasers usually does not exceed 5$\%$ (see, for example, \cite{41}). As calculations show, this is an acceptable value in order to neglect laser energy fluctuations in our scheme.

The different dependence of $\eta(\infty)$ and $\zeta(\infty)$ on wave functions $\Phi_I(t)$ and $\Phi_{II}(t)$  gives rise to the question of simultaneous fulfillment of equations (29), as was pointed out in Sec. IIB. There are many possibilities to overcome this difficulty by correctly choosing the parameters of the system, primarily for the $\Omega_2(t)$ laser field. We exemplary show in Fig. 4 that one can reduce the contribution of $\Phi_I(t)$ by delaying the laser pulse $\Omega_2(t)$ turning on with respect to the incoming photon pulses (see Fig. 3). In this case, the condition (23) is well satisfied for any time moment being insensitive to the spectral properties of the photon pulses. In the two other cases of $T_1=0.75 \mu s$ and $T_1=0.22 \mu s$ shown in Fig. 2, the conditions (29) are implemented in a similar way, leading, however, to smaller entanglement between the atoms due to inhomogeneous distribution of the sending atom population over the Zeeman states (see Table 1).

\subsection{Atomic state detection}
In our protocol, where the photon losses are considered negligible and the detectors have high efficiency, the entanglement succeeds with unit probability. In practice, the robustness of the proposed repeater has to be tested against inevitable noise in local operations and losses in quantum channels, although the latter is tiny within the fiber-channel attenuation length. The issues with photon losses occur at larger distances, this have been thoroughly discussed in our previous work \cite{27}. The effect of both photon loss and imperfect factors can be addressed by photon-heralding measurements, which, however, are challenging with our multiphoton protocol. Instead, we propose here to verify the successful storage of incoming photons in the second cavity by using a scheme for a reliable measurement of the ground Zeeman state populations of the second atom.

\begin{figure}[t] \rotatebox{0}{\includegraphics*
[scale =0.54]{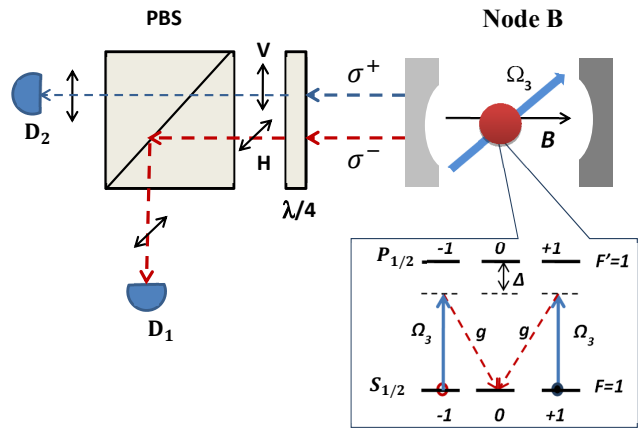}} \caption{(Color online) Setup for measuring the Zeeman states of the second atom at the node B. The states $|\bar{m}_F=-1\rangle_2$ and $|\bar{m}_F=+1\rangle_2$ of the receiving atom are converted into $\sigma^-$- and $\sigma^+$-polarized single photons, which are separately detected at the output ports D1 and D2, respectively, passing through a $\lambda/4$ plate and PBS.
\label{fig:fig5}}
\end{figure}

The schematic setup for atomic state detection is shown in Fig. 5, where the receiving atom is driven on the $S_{1/2}(F=1)\rightarrow P_{1/2}(F'=1)$ transition off-resonantly by a linearly polarized laser pulse $\Omega_3(t)$. The latter generates a $\sigma^+$- or $\sigma^-$-polarized cavity photon with probabilities $\beta_{\mp 1}^2$, respectively, dependent on the state of the atom. To detect these photons, we propose to use a two-mode cavity similar to that employed in \cite{42,43}. The first cavity mode is coupled to the atomic $F=1\rightarrow F'=2$ transition (Fig. 1) and supports the efficient storage of incoming photons, while the second mode with frequency of $\Omega_3(t)$ enhances the atomic state readout signal. A weak magnetic field $B$ is applied to reduce the local magnetic field fluctuations. The emitted photons pass through a quarter-wave plate ($\lambda/4$), which converts left- and right-polarized photons to be horizontally (H) and vertically (V) polarized, respectively. Upon reaching the interface of a polarizing beam splitter (PBS), the H polarized photon is reflected to the $D_1$ detector, while the V polarized photon goes straight to the $D_2$ detector. Thus, the states $|\bar{m}_F=+1\rangle_2$ and $|\bar{m}_F=-1\rangle_2$ of the receiving atom are selectively measured by single-photon detection. In the ideal case, the number $N_{\pm}$ of the events with detection of $\sigma^+$- and $\sigma^-$-polarized photons should be related as $N_{+}/N_{-}=\beta_{-1}^2/\beta_{1}^2$ according to Eq. (29). Deviations from this value will make it possible to determine the sources of imperfect transmission of photons. In this sense, the simplest situation to analyze is given in Fig. 2a (see also Table 1, $T_1=0.75\mu$s), where only the state $|\bar{m}_F=-1\rangle_2$ of the receiving atom is populated as the result of two-photon absorption. 
Correspondingly, the fidelity of the measured density matrix $\rho_2$ of the second atom with the calculated state (29) contains only one term  $F=|_2\langle\bar{m}_F=-1|\rho_2|\bar{m}_F=-1\rangle_2|^2$ to be analyzed statistically.
In the state $|\bar{m}_F = 0\rangle_2$, the atom does not interact with linearly polarized $\Omega_3(t)$ field, so no signal appears at PBS output. This state is then measured by applying a laser pulse tuned to the $S_{1/2}(F = 1,\bar{m}_F=0) \rightarrow P_{1/2}(F'=2,\bar{m}_F = 0)$ transition with a proper detuning, which triggers the emission of a $\sigma^+$- or $\sigma^-$-polarized photon at the frequency of the first cavity-mode on the transition $(F = 1,\bar{m}_F=-1)\rightarrow (F'=2,\bar{m}_F = 0)$ or $(F = 1,\bar{m}_F=1)\rightarrow (F'=2,\bar{m}_F = 0)$, respectively. The detection of this photon without discrimination of its polarization indicates that the atom is in the state $|\bar{m}_F = 0\rangle_2$. Furthermore, the entanglement created between the atoms according to Eq.(29) can also be verified using the scheme in Fig.5 at both nodes A and B. Upon converting the states of two atoms into the single photons and detecting the latter as the $\sigma^{\pm}$ events, one can make sure that a $\sigma^+(\sigma^-)$ event at the node A is always accompanied by the $\sigma^-(\sigma^+)$ event at the node B. If the atoms are in the states of $m_F=\bar{m}_F=0$, the absence of an output signal must be registered simultaneously at two nodes. The tomography data for the last case can be obtained by detecting again one photon from each node emitted from upper states $(F'=2,m_F = 0)$ and $(F'=2,\bar{m}_F = 0)$ of the atoms, as described above. Then, using only the successful attempts of two photon detection, the density matrix of the system can be reconstructed, from which the amount of entanglement can be obtained with high accuracy, if all conditions we use in our calculations are experimentally kept. Finally, we note an especially important possibility of functional integration of this scheme into QR structure at the stage of entanglement swapping between the entangled pairs.

\section{\protect\normalsize CONCLUSION}
We have proposed the first protocol for generating multidimensional entanglement between remote atoms in the space of atomic states of higher dimension than two. We have developed an experimentally feasible model for deterministic generation of three-dimensional entanglement between two distant atoms trapped in optical cavities and exchanging two photons. We have carried out the detailed analysis of a lossless evolution of the entanglement up to the creation of maximally entangled two-qutrit state of the atoms. In realistic conditions, the maximal entanglement has been calculated numerically to be 1.58, far above the two-dimensional limit of 1. The conditions for deterministic production of two photons from the sending cavity and their complete absorption by the receiving atom have been determined. The necessary condition in the latter case has been formulated explicitly by Eq. (17), which is a universal requirement to prohibit the reflection or leakage of incoming photons from a cavity. To test the robustness of our scheme against the imperfect local operations and photon losses in the fiber-quantum channel, we have proposed a technique, which is based on a reliable measuring of ground Zeeman states of the atoms and is simple to implement.

These results are the basis for performing an entanglement swapping operation between entangled pairs, which is under investigation.

\subsection*{Acknowledgments}

This work was supported by the RA Science Committee, in the frames of the research project 20TTAT-QTc004.
We acknowledge support from the EUR-EIPHI Graduate School (17-EURE-0002) and from the European Union's Horizon 2020 research and innovation program under the Marie Sklodowska-Curie grant agreement No. 765075 (LIMQUET).

\end{document}